\documentclass[twocolumn,prb,superscriptaddress]{revtex4-2}

\newcommand{\dn}{n'}  
\renewcommand{\dj}{J'}  

\newcommand{\e}{\mathrm e}

\usepackage{lipsum}
\usepackage{xcolor}
\usepackage{physics}
\usepackage{graphicx}
\usepackage{dcolumn}
\usepackage{bm}
\usepackage{color}
\usepackage[colorlinks,
            citecolor=blue,
            linkcolor=red,
            urlcolor=blue]{hyperref}

\definecolor{fgreen}{rgb}{0.13, 0.54, 0.13}

\begin{document}

\preprint{APS/123-QED}

\title{Terahertz Radiation from the Dyakonov-Shur Instability of Hydrodynamic Electrons in a Corbino Geometry}

\author{Jack H.\ Farrell}
\affiliation{Department of Physics, University of Colorado, Boulder CO 80309, USA}
\affiliation{Department of Physics, University of Toronto, 60 St. George Street, Toronto, Ontario, M5S 1A7, Canada}
\author{Nicolas Grisouard}%
\affiliation{Department of Physics, University of Toronto, 60 St. George Street, Toronto, Ontario, M5S 1A7, Canada}
\author{Thomas Scaffidi}
\affiliation{Department of Physics and Astronomy, University of California, Irvine, California 92697, USA}
\affiliation{Department of Physics, University of Toronto, 60 St. George Street, Toronto, Ontario, M5S 1A7, Canada}

\renewcommand{\dv}{v'}  
\date{\today}

\begin{abstract}
Hydrodynamic electrons flowing through a two-dimensional channel are predicted to undergo a plasma instability above a critical drift velocity.
This Dyakonov-Shur (DS) instability terminates as a coherent nonlinear oscillator, which shows promise as a source of radiation that could fill the so-called TeraHertz gap. 
In this work, we study radial flow in a Corbino disk, and demonstrate how the DS instability is substantially enhanced in this geometry, both in terms of a lower critical drift velocity and a higher generated power. Interestingly, hydrodynamic electron flows were recently reported in a graphene sample of this geometry, and our results are therefore directly relevant to current efforts to detect this experimentally elusive phenomenon.
 The analysis is based on a hydrodynamic approach and features both linearized calculations as well as full numerical simulations of the Navier-Stokes equation. 

\end{abstract}

\maketitle

\section{Introduction}\label{sec:introduction}

 The TeraHertz (THz) gap refers to a frequency range of the electromagnetic spectrum between 0.1 and 10 THz, which has been underutilized due to a lack of practical sources and detectors~\cite{Dhillon_2017}.
Closing this gap would pave the way for numerous applications in the industry (e.g. telecommunications, imaging) as well as in fundamental science (e.g. solid-state spectroscopy, astronomy). 
Consequently, the search for practical sources of TeraHertz radiation has been an ongoing area of research for several decades~\cite{Dhillon_2017,Lewis_2014}.

A promising solution, first proposed by Dyakonov and Shur~\cite{Dyakonov1993}, is to utilize plasma oscillations in a two-dimensional electron system.
These oscillations generate radiation in the THz range if the typical length scale of the sample is on the order of 1 $\mu$m.
Although first proposed in the context of GaAs heterostructures, two-dimensional materials like graphene~\cite{Koppens2014,WANG2018107} and van der Waals heterostructures~\cite{Geim2013,Mounet2018} have since emerged as other promising candidates for realizing this proposal, thanks in part to their extremely long momentum-relaxing mean free path.

Interestingly, this quasi-conservation of momentum, combined with strong electron-electron scattering, is also responsible for a hydrodynamic regime of transport that has attracted tremendous attention over the last few years~\cite{gurzhi1963minimum,gurzhi1968hydrodynamic,PhysRevB.21.3279,PhysRevLett.52.368,gurzhi1995electron,PhysRevB.49.5038,PhysRevB.51.13389,PhysRevLett.52.368,PhysRevLett.71.2465,PhysRevLett.77.1143,Spivak20062071,PhysRevLett.106.256804,PhysRevB.92.165433,PhysRevLett.117.166601,PhysRevLett.113.235901,levitov,PhysRevB.92.165433,PhysRevB.93.125410,PhysRevB.94.125427,2016arXiv161209239G,PhysRevB.95.115425,PhysRevB.95.121301,Bandurin1055,Crossno1058,Moll1061,Narozhny2017,Guo3068,Kumar2017,2017_PRL_ScaffidiNSMM,PhysRevB.97.045105,PhysRevB.97.121404,PhysRevLett.121.176805,PhysRevB.98.165412,PhysRevB.97.121405,Gooth2018,Braem2018,Berdyugin2019,Sulpizio2019,NAROZHNY2019167979,Shavit2019,Ku2020,LEVCHENKO2020168218,PhysRevB.100.245305,Jenkins2020,Keser2021,Gupta2021,Krebs2021,Hong2020, Sukhachov}. 
In this regime, Ohm's law is supplanted by the Navier-Stokes (NS) equation, and viscosity becomes a dominant contribution to the resistance.
The linear regime of the Navier-Stokes equation, in which the convection term can be neglected, has by now been studied extensively in electronic systems. This has led to numerous experimental discoveries, like the detection of current vortices~\cite{Bandurin1055}, the realization of superballistic flows~\cite{Kumar2017}, the measurement of the Hall viscosity in graphene~\cite{Berdyugin2019}, and the direct visualization of a Poiseuille flow of electrons~\cite{2019_Nature_Sulpizioetal}.
By contrast, the study of hydrodynamic electron flows in the non-linear regime of NS, with the prospect of realizing phenomena like pre-turbulence, is still in its infancy~\cite{PhysRevLett.121.236602}.

In this context, the Dyakonov-Shur instability can be understood as a non-trivial manifestation of the non-linear regime of the Navier-Stokes equation~\cite{Dyakonov1993,2021_APL_MendlPL,2021PhRvB.104o5440C}. 
This instability arises for a field effect transistor (FET) with asymmetric boundary conditions at the source and drain, in the presence of a bias DC current above a certain threshold.
Dyakonov and Shur used linearized equations to prove the presence of an instability towards oscillations in the THz range.
Further, recent numerical simulations~\cite{2021_APL_MendlPL}{} have observed the endpoint of this instability to be a coherent linear oscillator, making the phenomenon even more promising as a radiation source.
However, despite extensive work on the topic~\cite{Dyakonov1993,doi:10.1063/1.1391395,doi:10.1063/1.2042547,2021_APL_MendlPL,Sydoruk2010,2021PhRvB.104o5440C,PhysRevApplied.10.024037,PhysRevB.99.075410,Li2019,Khavronin2020}, an experimental detection of the DS instability is still lacking.

\begin{figure}[t!]
    \centering
    \includegraphics[width=0.7\columnwidth]{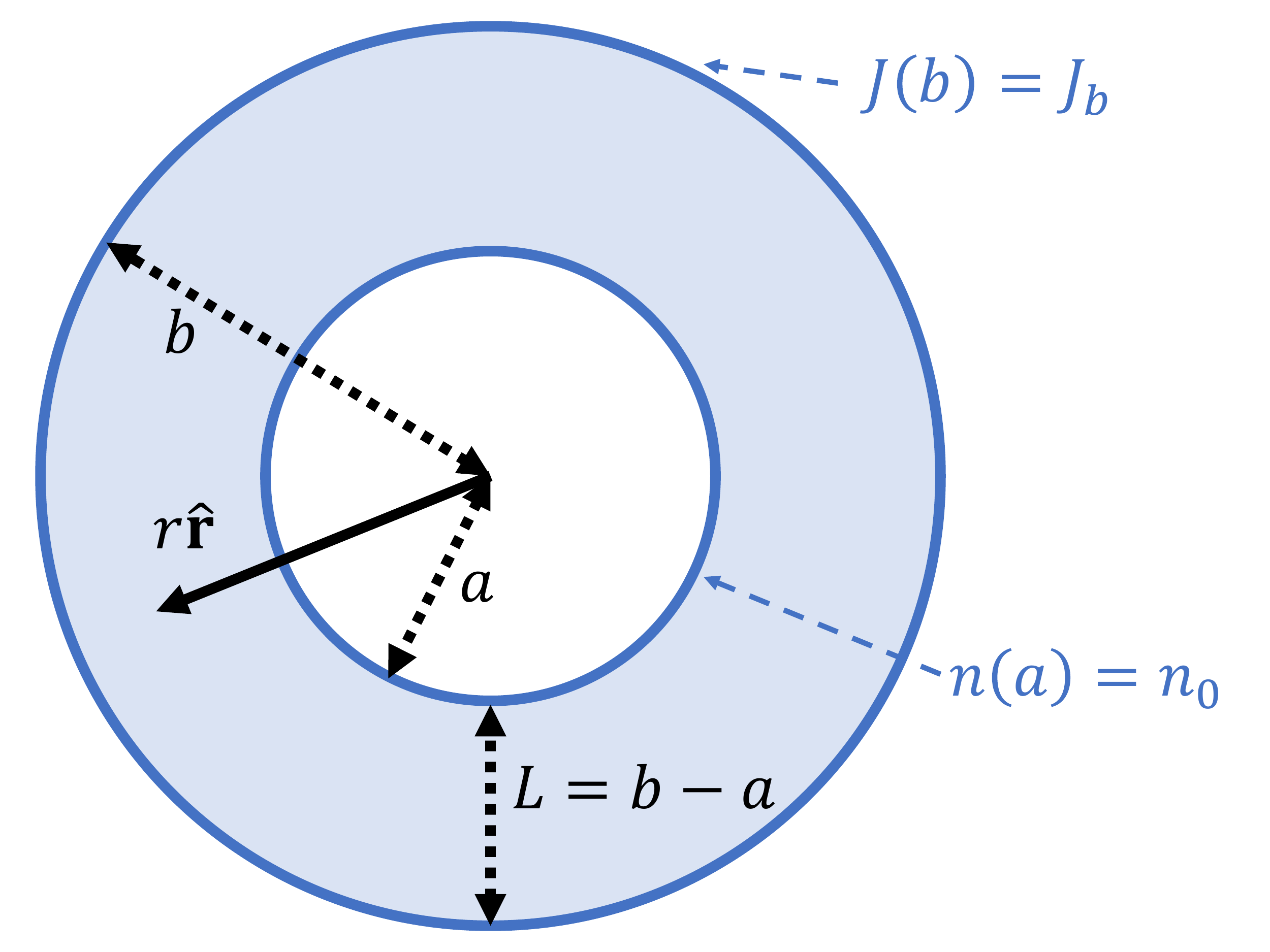}
    \caption{We model radial flow in a Corbino disk $a<r<b$. The difference $L = b - a$ is the analog of the channel length in a rectangular device. Boundary conditions are $n(a)=n_0$, $J(b)=J_b= n_0 v_b$.}
    \label{fig:SketchCorbino}
\end{figure}

In this work, we show how to enhance the plasma instability by an advantageous choice geometry.
As shown in Fig.~\ref{fig:SketchCorbino}, we study radial flow in a Corbino disk, with asymmetric boundary conditions between the inner and outer contact.
A strong dependence on geometry can generically be expected in the hydrodynamic regime due to the non-local relation between current and electric field.
In fact, the performance of plasma oscillations for the generation and detection of radiation has been found to depend strongly on the geometry in previous work~\cite{Sydoruk2010,Li2019,Khavronin2020}.

We can expect the annular geometry to enhance the instability based on symmetry grounds alone. 
In a rectangular channel, the instability is driven by the asymmetric boundary conditions that differ at the two ends of the channel, which allow plasma waves to be amplified upon reflection from the boundaries.  In contrast, symmetric boundary conditions do not lead to an instability in the rectangular channel.  In the Corbino case, there is an additional geometric asymmetry between the two ends of the channel, i.e., the inner and outer circular boundaries.  We thus expect that combining asymmetric boundary conditions with the geometric asymmetry will lead to a stronger instability. 

Our work is also motivated by a recent experiment which observed hydrodynamic electrons flowing in a Corbino geometry~\cite{Kumar2021}.
A peculiar feature of this geometry is the absence of lateral walls, i.e. walls which are parallel to the flow.
This means the viscous term generates no bulk resistance for a radial DC current in an annulus, as predicted theoretically in Ref.~\cite{Shavit2019} (See also Ref.~\cite{Kumar2021} for a recent experimental demonstration of this effect).
This is to be contrasted to a rectangular channel of finite width $W$, for which scattering at the lateral boundaries leads to a partial slip boundary condition and ultimately to resistance.

The article is organized as follows. In Section II, we describe our hydrodynamic model for a Corbino FET, which includes the effects of two types of dissipation: viscosity and momentum relaxation. In Section III.A, we study perturbations around a simple DC bias current to find the growth rate and frequency of the dominant quasinormal mode in the linearized regime. In Section III.B, we present results of a full simulation of the Navier-Stokes equation. We show that the instability saturates as a coherent non-linear oscillator, and that increasing the annularity of the system enhances the DS instability.
We finish with a conclusion in Section IV.

\section{Model\protect}\label{sec:model}

We model a two-dimensional electron gas with a quadratic dispersion relation parameterized by an electron mass $m$, in an annular geometry defined by $a < r < b$ (Fig.~\ref{fig:SketchCorbino}). Electrons are described by their number and momentum densities $n(r, t)$ and $J(r,t)\equiv n(r,t)v(r,t)$, respectively, with $v(r,t)$ the velocity field.  We assume that the flow is purely radial, both in directions and spatial dependence. The assumption of radially symmetric flow, which leads to an effectively one-dimensional model, is justified by the azimuthal invariance of the boundary conditions~\eqref{eq:BCs}. Additionally, electrons in this regime will experience both viscous dissipation and momentum relaxation.  To quantify these effects, let $\eta$ be the coefficient of dynamic viscosity and $\gamma \equiv 1/\tau$ be the momentum relaxation rate.  Finally, we also assume an external static electric field $\vb E \equiv E(r) \vu r$ and a thermodynamic pressure $P(n)$, giving the Navier-Stokes equation
\begin{equation}
\begin{split}
\pdv{J}{t}+\frac{1}{r}\pdv{r}(\frac{rJ^2}{n}) + \frac{1}{m}\pdv{P}{r} &-  \frac{\eta}{m}\left(\laplacian-\frac{1}{r^2}\right) \frac{J}{n}
\\&= -\gamma J - \frac{en}{m}E(r), \label{eq:momentumEquation}
\end{split}
\end{equation}
along with the continuity equation
\begin{equation}
\pdv{n}{t} + \frac{1}{r}\pdv{r}(rJ) = 0. \label{eq:massEquation}
\end{equation}
Here, $\laplacian \equiv \partial_r^2 +r^{-1}\partial_r$ 
stands for the Laplacian of an axisymmetric scalar function.

\begin{figure}[t!]
    \centering
    \includegraphics{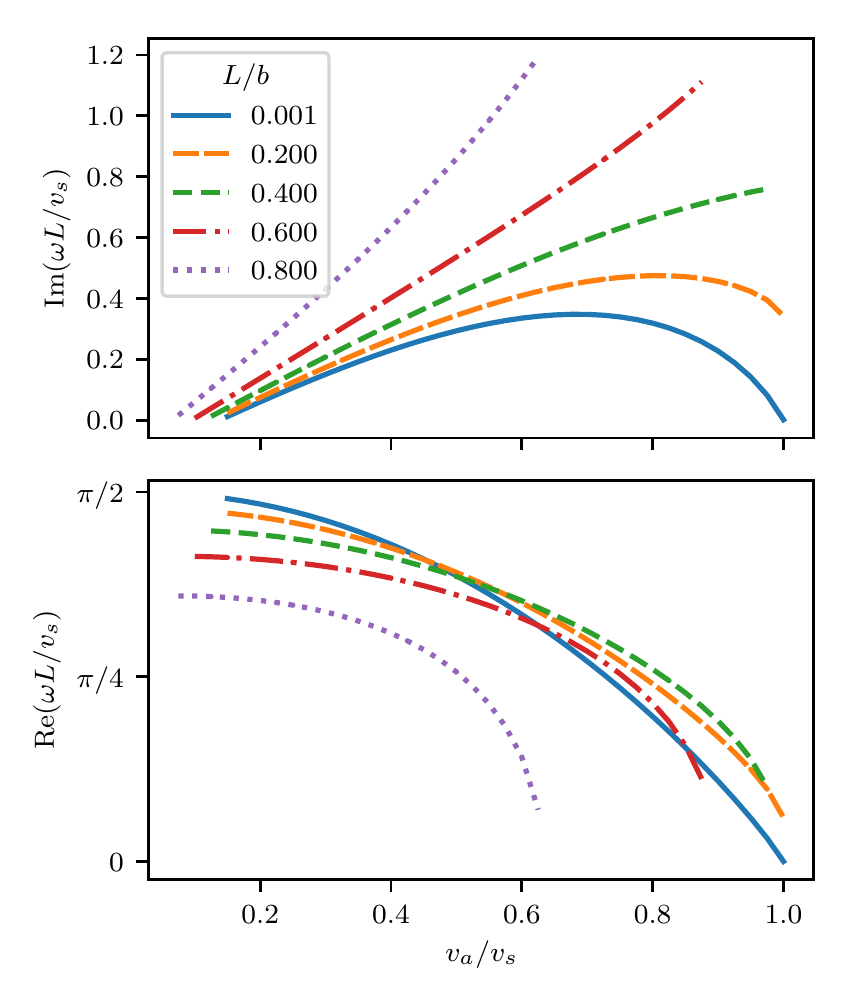}
    \caption{Growth rate (imaginary part, top) and oscillation frequency (real part, bottom) of quasinormal modes as a function of the drift velocity $v_a$ for increasingly annular devices (\textit{i.e.} increasing $L/b$).  Parameters are $\tilde\eta = 0.03$ and $\tilde\gamma = 0.2$.}
    \label{fig:linear}
\end{figure}

We impose so-called \textit{Dyakonov-Shur Boundary Conditions} \cite{Dyakonov1993}, fixing the density at the inner contact ($r=a$) and the current at the outer contact ($r=b$), namely,
\begin{subequations}
	\label{eq:BCs}
	\begin{align}
	n(a) &= n_0, \label{eq:bc1} \\
	J(b) &= J_b, \label{eq:bc2}
	\end{align}
\end{subequations}
where $J_b =  n_0 v_b = n_0 v_a b / a$ is the imposed current at the outer radius, and $v_a$ is the velocity at the inner radius, as we will see below. 
For the pressure, following \citet{2021_APL_MendlPL}, we use a simplified equation of state corresponding to an ideal gas, namely, $P(n) = m v_s^2 n$, where $v_s$ is the speed of sound and is assumed to be constant.
In quoting results from the simulations, it will be useful to work in terms of the following non-dimensional parameters $v_a$, $\eta$, $\gamma$, that is, 
\begin{equation}
    \tilde{v}_a\equiv \frac{v_a}{v_s},\quad \tilde{\eta} \equiv \frac{\eta}{n_0 m v_s L}\quad \text{and}\quad \tilde{\gamma} \equiv \frac{\gamma L}{v_s}.
\end{equation}

We study perturbations around a steady state solution to Eqs.~\eqref{eq:momentumEquation} and \eqref{eq:massEquation}. For the sake of simplicity, we aim to study a steady state with uniform density, $n_0(r) = n_0$. The current simply follows from the continuity equation: 
\begin{equation}
    \label{eq:steadystate}
    J_0(r) = \frac{J_b b}{r} 
\end{equation}
We then choose the external static electric field $E(r)$ so as to ensure this solution is a steady state, which leads to
\begin{equation}
\frac{-e}{m} E(r)= \gamma \frac{J_b b}{n_0 r} - \frac{J_b^2 b^2}{n_0^2 r^3}.
\end{equation}

Additionally, to clarify the relationship with the rectangular geometry, we define $L \equiv b - a$ as the channel length.  Then, as we let $L / b \rightarrow 0$, Eqs.~\eqref{eq:momentumEquation} and \eqref{eq:massEquation} reduce to those for a straight channel.  As such, we use the quantity $L/b$ as a measure of the difference between the rectangular geometry and our Corbino disk geometry. We will show that increasing $L/b$ enhances the instability compared to the rectangular case ($L/b=0$) reported in \cite{2021_APL_MendlPL}. Finally, we note that some care must be taken when comparing the critical drift velocity between a rectangular device and an annular one, since the drift velocity is not uniform in the latter case. In order to be conservative, when defining the critical drift velocity, we use the drift velocity at the inner contact of the annulus, where it takes its maximal value.

\begin{figure}[t!]
    \centering
    \includegraphics{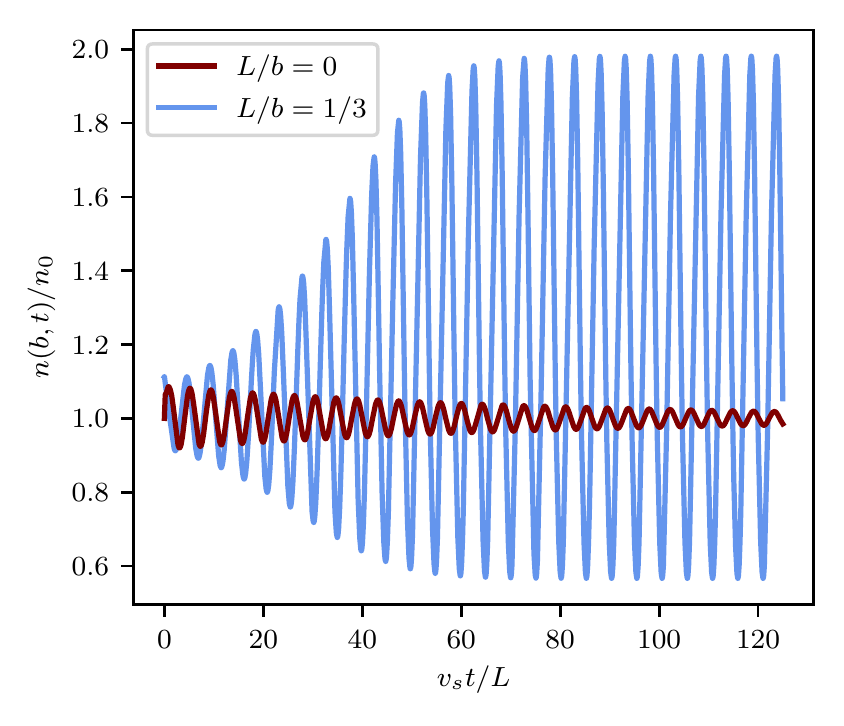}
    \caption{Time series of $n(b,t)$, the density evaluated at the end of the channel, for a rectangular geometry ($L/b\rightarrow 0$, burgundy trace) and for a Corbino geometry ($L/b=1/3$, blue trace).  The parameters were $\tilde v_a = 0.25$, $\tilde\eta = 0.1$, $\tilde\gamma = 0.16$.}
    \label{fig:oscillation}
\end{figure}

\section{Results}
\subsection{Linear analysis}
First, we study the oscillations around the steady state by considering perturbations of the form:
\begin{align}
\begin{split}
    n &= n_0 + \dn  \e^{- i \omega t}, \\
    J &= J_0(r) + \dj  \e ^ {- i \omega t},
\end{split}
\end{align}
where the real part of $\omega$ gives the oscillation frequency and the imaginary part gives a \textit{growth rate}. With our convention, $\Im \omega > 0$ means that the perturbation grows.  Neglecting higher order terms in $\dj$ and $\dn$ gives the following set of linear equations:
\begin{subequations}
    \label{eq:annular_linear_no_visc}
\begin{align}
    \label{eq:vel}
     \pdv{\dj}{t} + \frac 1 r \pdv{r}(r \frac{2 J_0 \dj}{n_0} - r\frac{J_0^2 \dn}{n_0^2}) + v_s^2 \pdv{\dn}{r} 
     \nonumber\\- \frac{\eta}{m}\left(\nabla^2 - \frac{1}{r^2} \right)\left( \frac{\dj}{n_0}- \frac{J_0\dn}{n_0^2} \right) + \frac{v_b^2 b^2}{r^3} \dn  &= -\gamma \dj + \gamma \frac{v_bb}{r}\dn,\\
     \pdv{\dn}{t} + \frac 1 r\pdv{r}(r \dj) &= 0.
\end{align}
\end{subequations}
We find the quasinormal modes of Eqs.~(\ref{eq:annular_linear_no_visc}) numerically with an Arnoldi scheme and show the growth rate and frequency of the dominant one in Fig.~\ref{fig:linear}.

For a perfectly rectangular device ($L/b \rightarrow 0$), the growth rate increases with $v_a$ at low velocity and vanishes as $v_a$ approaches the speed of sound.  In contrast, for $L/b = 0.6$, for example, the instability persists as $v_a$ increases to the speed of sound; indeed, the growth rate for nonzero $L/b$ is systematically greater than that of the rectangular case, and is an increasing function of $L/b$.

We have thus found that the growth rate increases with $L/b$, which also means that the DS instability starts at a lower critical velocity.
We have also checked that, for a fixed $v_a$, the instability survives for larger values of the dissipation coefficients $\tilde\eta$ and $\tilde\gamma$.

\subsection{Simulations}
Moving beyond the linearized regime, we use finite volume methods to perform a real-time simulation of Eqs.~\eqref{eq:momentumEquation} and \eqref{eq:massEquation} (see Appendix A for more details about the numerical method). 
Like in the rectangular geometry \cite{2021_APL_MendlPL}, we observe that the initial growth eventually saturates due to dissipation, and converges to coherent oscillations for which neither the frequency nor the amplitude depend on the initial conditions (see Fig.~\ref{fig:oscillation}). 

\begin{figure}[t!]
    \centering
    \includegraphics[width=0.85\columnwidth]{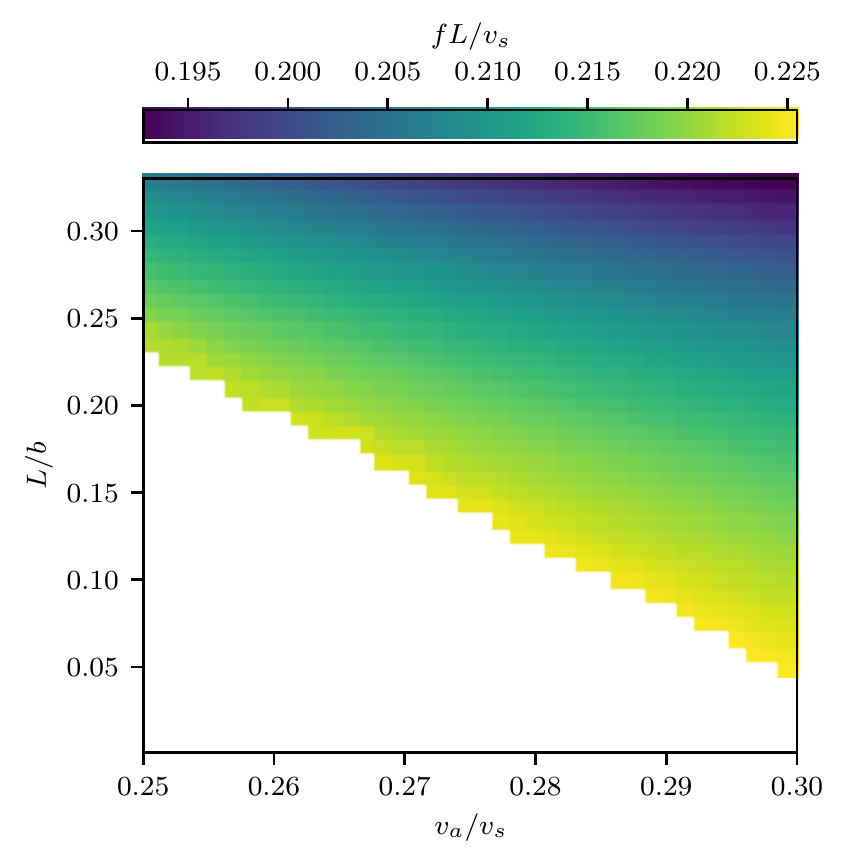}
    \caption{Frequency of nonlinear oscillator as a function of $v_a$ and $L/b$ at fixed $\tilde{\eta} = 0.1$ and $\tilde \gamma =  0.2$.  Regions of white denote no instability.  We find that the critical $v_a$ for instability decreases as $L/b$ is increased. }
    \label{fig:existence}
\end{figure}

\paragraph{Existence of Instability}
Our simulations verify that the Corbino disk geometry enhances the instability compared to the rectangular geometry.  As an illustrative example,  Fig.~\ref{fig:oscillation} shows time series of the oscillation of $n(b,t)$, the density evaluated at the end of the channel, in the case $(\tilde v_a,\ \tilde \eta,\ \tilde\gamma) = (0.25,\ 0.1,\ 0.16)$. In the rectangular geometry ($L/b=0$), we find a decaying oscillation, but for $L/b=1/3$, the solution grows until it saturates as a coherent oscillator.  This behavior is general: a higher $L/b$ allows the instability to exist in previously stable regions of the parameter space.
Additionally, Fig.~\ref{fig:existence} shows how the critical drift velocity decays approximately linearly with $L/b$, and how the frequency of the final oscillations mildly depends on $L/b$.

\paragraph{Radiated Power}
\begin{figure}[t!]
    \centering
    \includegraphics[width=\columnwidth]{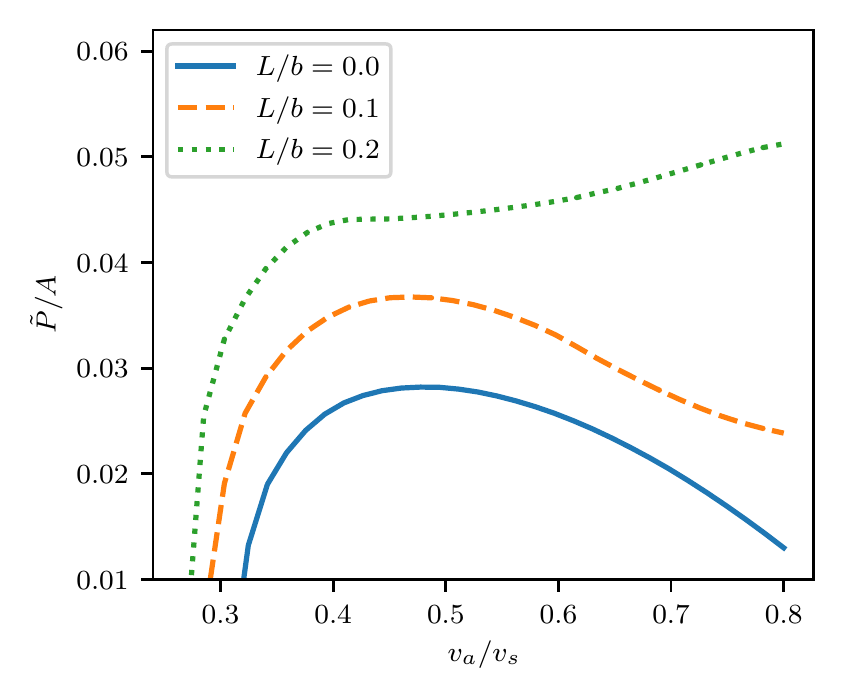}
    \caption{Dimensionless power per unit area,  $\tilde{P}/A=P L \epsilon_0 \epsilon_{zz} / d e^2 n_0^2 v_s$, as a function of bias velocity for three values of $L/b$.  Making the device more annular (increasing $L/b$) gives a larger radiated power. Other parameters are $\tilde \eta = 0.1, \tilde \gamma = 0.2$.}
    \label{fig:numerical_power}
\end{figure}
To go further in the comparison between annular and rectangular devices, it helps to estimate the power that the oscillator could radiate. If we consider one device with $L \approx 1\ \mu\text{m}$, the actual power radiated could be approximated using the formula for dipole radiation~\cite{jackson}:
\begin{align}
    P = \frac{\omega^4 Z_0}{12 \pi c^2} |\vb{p}|^2,
\end{align}
with $Z_0=\sqrt{\mu_0/\epsilon_0}$ the impedance of free space, $|\vb p| = Q d$ the dipole moment, and $Q$ the maximum charge in the device.  For $\mu\text{m}$-scale devices, such a calculation yields a power on the order of $p\text{W}$, which is small compared to typical $m\text{W}$-scale measuring devices.

However, the true electrostatic energy stored in the device is much larger than that available for dipole radiation.  Accordingly, for realistic applications, devices like the one proposed in this letter are routinely connected to amplification mechanisms such as \emph{antennae}.  For the purposes of this letter, it is best to give an upper bound on the available power in a way that is agnostic to the details of the amplification mechanism: we assume a \emph{perfect} amplifier that can access \emph{all} the electrostatic energy stored in the device.   Following Ref.~\cite{2021_APL_MendlPL}, we approximate the FET as a capacitor undergoing charge-discharge cycles, leading to an upper estimate of the power given by
\begin{equation}
\label{eq:rectPower}
P = \frac{1}{T}\left| \frac{1}{T}\int_t^{t+T} \frac{1}{2C}Q^2(t)\cos(\frac{2 \pi t}{T}) \mathrm{d}t\right|,
\end{equation}
where $C = \epsilon_0\epsilon_{zz}A / d$ is the geometric capacitance and  $A$ is the area of the capacitor (which is given by $A = LW$ in the rectangular case and $A=\pi(b^2-a^2)$ for the annulus).  The charge $Q(t)$ is found by integrating the charge density $n$,
\begin{equation}
\label{eq:circCharge}
Q(t) = -2 \pi e\int_a^b r n(r,t)\mathrm{d}r.
\end{equation}

In Fig.~\ref{fig:numerical_power}, we show the radiated power calculated from Eq.~(\ref{eq:rectPower}).
We notice that the radiated power per unit area increases for more annular devices (higher $L/b$).
 Further, for $L/b=0.2$, $P$ shows a monotonic increase with bias velocity, in contrast to the broad maximum shown in the rectangular case.
 
 \paragraph{Amplifying the power}
Before concluding, it is worth commenting on a few practical points that will be relevant for realizations of the DS instability in Corbino FETs.  For one, as we have already established, true amplifiers or antennae will \emph{not} access the whole electrostatic energy stored inside our device, as we have assumed in the estimates above.  The inefficiencies will be experiment-dependent and should be calculated in-situ.  However, such an amplifier or antenna would also lead to radiative damping due to back action from the emitted radiation, which would tend to decrease the instability growth rate.  Realistic calculations or simulations of devices connected to particular antenna patterns will be needed for accurate predictions that can precisely inform experiments.

\section{Conclusion}

The annulus is a particularly interesting configuration to study hydrodynamic transport. In fact, predictions of vanishing bulk resistance and non-trivial magnetoresistance~\cite{Shavit2019} in this geometry were recently confirmed in experiments~\cite{Kumar2021}.
In this work, we have shown that the non-trivial properties of this geometry also extend to the case of non-linear transport.
We studied the DS instability in this annular geometry by combinining a perturbative analysis of the linearized regime and a full numerical simulation of the non-linear NS equations.
Compared to the straight channel, we found a strongly increased tendency towards the DS instability and a larger radiated THz power, which should make it easier to observe this phenomenon in experiments.

An aspect left for future work is a careful study of the spatial features of the plasma oscillations, and whether they could be detected with the imaging probes developed in Ref.~\cite{Kumar2021}.
As explained in \citet{Stern2021}, the Corbino geometry can also be understood as a limit of a more general ``wormhole'' geometry, which makes it possible to completely suppress the Landauer-Sharvin resistance.
The fate of the DS instability in these and other types of geometries is an interesting research direction. 

Whereas this work assumed a priori a hydrodynamic description, the regime observed in experiments is usually in-between ballistic and hydrodynamic.
In fact, in order to reach THz frequencies, $L$ would typically need to be of the order of 1~$\mu m$, which is only a few times larger than the typical electron-electron mean free paths measured in graphene in the relevant temperature window.
This does not by itself preclude the possibility of realizing the DS instability in graphene, however, since it was shown to persist even in the ballistic regime~\cite{doi:10.1063/1.1391395,doi:10.1063/1.2042547,PhysRevB.99.075410}. A kinetic study of the fate of the DS instability across the ballistic to hydrodynamic crossover~\cite{Mendl2018ballistic} in a Corbino geometry is left for future work.
Further, since graphene is one of the main candidate materials to realize hydrodynamic instabilities, it will be important to generalize the current calculation to the case of relativistic materials, for which higher-order non-linear terms appear in Euler's equation~\cite{PhysRevB.88.205426,PhysRevB.88.245444}, and for which temperature variations across the sample cannot be neglected~\cite{Sukhachov,PhysRevB.105.125302}.

\begin{acknowledgements}
We would like to acknowledge helpful discussions with Shahal Ilani and Andrew Lucas. This work was made possible by the facilities of the Shared Hierarchical Academic Research Computing Network (SHARCNET: \href{www.sharcnet.ca}{www.sharcnet.ca}) and Compute/Calcul Canada. JF was supported by a Natural Sciences and Engineering Research Council of Canada (NSERC) Undergraduate Summer Research Award. NG acknowledges financial support from NSERC's Discovery Grant program [RGPIN-2015-03684]. TS acknowledges the support of NSERC, in particular the Discovery Grant [RGPIN-2020-05842], the Accelerator Supplement [RGPAS-2020-00060], and the Discovery Launch Supplement [DGECR-2020-00222].
\end{acknowledgements}

\bibliography{references}

\newpage\clearpage

\appendix

\section{Numerical Methods}
Momentarily assuming $\eta = \gamma = 0$, the equations we study can be written as:

\begin{equation}
\label{eq:conservationLaw}
\pdv{t}
\begin{pmatrix}
n \\
J
\end{pmatrix}
+ \pdv{r}
\begin{pmatrix}
J \\
J^2/n + v_s^2n
\end{pmatrix}
=
\begin{pmatrix}
-J/r \\
-J^2/(nr)
\end{pmatrix}.
\end{equation}
The left-hand-side of this equation is equivalent to the \textit{isothermal equations} from gas dynamics in one spatial dimension.  The right-hand-side represents additional source terms from the Corbino geometry (``geometric source terms").

For the left-hand-side of~(\ref{eq:conservationLaw}), which is called the \textit{conservation part} of the equation, we use finite volume methods.  In particular, we use Roe's Approximate Riemann Solver with a slope limiter (we choose the \texttt{minmod} limiter) to make a high-resolution method.  The simulation has a spatial step width $h=1/50$ and a temporal step width $k = 0.0005$.

The viscous term is incorporated into the above \textit{conservation step} of our solution by using the centered finite-differences approximations to the derivatives.  For the vector Laplacian in polar coordinates, this looks like:
\begin{equation}
\frac{\left(r + \tfrac{h}{2} \right)}{rh}\frac{q(r+h)-q(r)}{h} - \frac{\left(r-\tfrac{h}{2}\right)}{rh}\frac{q(r)-q(r-h)}{h} -\frac{q}{r^2},
\end{equation}
where $q \equiv J/n$.  Note that the last term, which has no derivatives, could just as well have been incorporated into the other step of the simulation, which is described in the next paragraph.

Finally, for the relaxation term and the geometric source terms (right-hand-side of Eq.~(\ref{eq:conservationLaw})), we use an operator splitting method called \textit{Strang Splitting}.


\end{document}